\documentclass[12pt]{jpconf} 
\usepackage{graphicx}
\begin{document}
\title{Analysis of Dynamic Multiplicity Fluctuations at PHOBOS} 
\author{
\centerline{Zhengwei Chai$^2$ for the PHOBOS Collaboration}
%
%
B B Back$^1$,
M D Baker$^2$,
M Ballintijn$^4$,
D S Barton$^2$,
R R Betts$^6$,
A A Bickley$^7$,
R Bindel$^7$,
A Budzanowski$^3$,
W Busza$^4$,
A Carroll$^2$,
Z Chai$^2$,
M P Decowski$^4$,
E Garc\'{\i}a$^6$,
N George$^{1,2}$,
K Gulbrandsen$^4$,
S Gushue$^2$,
C Halliwell$^6$,
J Hamblen$^8$,
G A Heintzelman$^2$,
C Henderson$^4$,
D J Hofman$^6$,
R S Hollis$^6$,
R Ho\l y\'{n}ski$^3$,
B Holzman$^2$,
A Iordanova$^6$,
E Johnson$^8$,
J L Kane$^4$,
J Katzy$^{4,6}$,
N Khan$^8$,
W Kucewicz$^6$,
P Kulinich$^4$,
C M Kuo$^5$,
W T Lin$^5$,
S Manly$^8$,
D McLeod$^6$,
A C Mignerey$^7$,
R Nouicer$^6$,
A Olszewski$^3$,
R Pak$^2$,
I C Park$^8$,
H Pernegger$^4$,
C Reed$^4$,
L P Remsberg$^2$,
M Reuter$^6$,
C Roland$^4$,
G Roland$^4$,
L Rosenberg$^4$,
J Sagerer$^6$,
P Sarin$^4$,
P Sawicki$^3$,
W Skulski$^8$,
P Steinberg$^2$,
G S F Stephans$^4$,
A Sukhanov$^2$,
J -L Tang$^5$,
A Trzupek$^3$,
C Vale$^4$,
G J van~Nieuwenhuizen$^4$,
R Verdier$^4$,
F L H Wolfs$^8$,
B Wosiek$^3$,
K Wo\'{z}niak$^3$,
A H Wuosmaa$^1$,
B Wys\l ouch$^4$}
%
%
%
%
\address{
$^1$~Argonne National Laboratory,  Argonne, IL 60439-4843, USA   \\ 
$^2$~Brookhaven National Laboratory,  Upton, NY 11973-5000, USA   \\ 
$^3$~Institute of Nuclear Physics, Krak\'{o}w, Poland,    \\ 
$^4$~Massachusetts Institute of Technology,  Cambridge, MA 02139-4307, USA   \\ 
$^5$~National Central University, Chung-Li, Taiwan   \\ 
$^6$~University of Illinois at Chicago, Chicago, IL 60607-7059, USA   \\ 
$^7$~University of Maryland,  College Park, MD 20742, USA   \\ 
$^8$~University of Rochester, Rochester, NY 14627, USA  
}

\begin{abstract}
This paper presents the analysis of the dynamic fluctuations in the 
inclusive charged particle multiplicity measured by PHOBOS for Au+Au 
collisions at $\sqrt{s_{NN}}=200$GeV within the pseudo-rapidity
range of $-3<\eta<3$. First the definition of the fluctuations observables used in 
this analysis is presented, together with the discussion of their physics
meaning. Then the procedure for the extraction of dynamic fluctuations is 
described. Some preliminary results are included to illustrate the
correlation features of the fluctuation observable. New dynamic
fluctuations results will be available in a later publication.
\end{abstract}





One of the important goals of the experiments at the Relativistic Heavy Ion 
Collider (RHIC) is to search for the evidence of the quark-gluon plasma(QGP)
/hadron gas(HG) phase transition. 
Event-by-event fluctuations might be a useful signature of phase 
transition evidence, since they could be significantly altered if a
phase transition occurs shortly after the collision\cite{phase_fluc} due to
the the large difference between the degrees of freedom of the two phases.
Since fluctuation observables are intrinsically related to particle 
correlations,  the study of fluctuations could also provide helpful
insights on the mechanisms of particle production in heavy ion collisions.


In this analysis, we measure the inclusive charged particle multiplicity 
fluctuations as a function of the separation and bin size in
pseudo-rapidity($\eta$)
space in Au+Au collisions with an energy of $\sqrt{s_{NN}}=200$GeV. 
Since the event-by-event particle multiplicity is mainly determined by
the the number of participating nucleons ($N_{part}$) in the collision,
a significant amount of the multiplicity fluctuations is due to the $N_{part}$
variation,  which is just a geometry effect. To study the dynamic fluctuations which are
related to intrinsic particle production mechanisms, the $N_{part}$
fluctuations must be removed from the measurement of multiplicity
fluctuations.

For the study of multiplicity fluctuations, we use an event-by-event
observable $C$, which is defined as
\begin{equation}
C = \frac{N_{1}-N_{2}}{\sqrt{N_{1}+N_{2}}},
\end{equation}
where $N_1$ and $N_2$ are the multiplicities in a pair of $\eta$ bins 
with the same bin size and symmetric with respect to $\eta=0$. The width of the $C$
distributions ($\sigma(C)$) is used as our fluctuations observable. Because 
the difference of the two multiplicities is used in the definition of $C$
and the two multiplicities change in the same direction when $N_{part}$
varies from event to event, the $N_{part}$ fluctuations are suppressed in 
the measured $\sigma(C)$. Also because of event-by-event 
normalization factor $\sqrt{N_{1}+N_{2}}$ in the denominator, $\sigma(C)=1$
for independent particles. Thus non-1 $\sigma(C)$ indicates non-0 dynamic
fluctuations in addition to the statistical fluctuations. We can decompose 
the multiplicity fluctuations  $\sigma^2(C)$  into two parts: statistical
fluctuations ($\sigma^2_{stat}$) and dynamic fluctuations ($\sigma^2_{dyn}$).
While the statistical fluctuations are due to the finite multiplicity,
the dynamic fluctuations are related to the intrinsic correlations
in the particle production. The long range (particles from different bins)
correlations and the short range (particles usually in the same bin) have
different effect in $\sigma^2(C)$ and different signs of the dynamic fluctuations.  

To understand what effect the long range correlations 
have on $\sigma^2(C)$, we can  use the approximate expression for $\sigma^2(C)$: 
$\sigma^2(C) \approx \frac{\sigma^2(N_1)+\sigma^2(N_2)-2\rho\sigma(N_1)\sigma(N_2)}{<N_1+N_2>}
\approx 1-\rho$ where $\rho$ is the correlation coefficient between $N_1$
and $N_2$, assuming the multiplicity $N$ follows Poisson distribution
($\sigma(N)=\sqrt{N}$). With such approximation, $\sigma^2_{dyn} \approx
-\rho$,  positive long range correlations lead to negative dynamic fluctuations
and reduce $\sigma^2(C)$ from 1. The short range correlations can be treated 
as a cluster effect and related to the UA5 cluster multiplicity
study\cite{ua5}. Assuming the particles are created in cluster with
multiplicity $k$, and the multiplicity in the investigated bin is increased
by a factor of $k$, $C=\frac{N_1-N_2}{\sqrt{N_1+N_2}}$ is increased to
$\sqrt{k}C=\frac{kN_1-kN_2}{\sqrt{kN_1+kN_2}}$ and $\sigma(C)$ is changed to
$\sqrt{k}\sigma(C)$ correspondingly. We then have $\sigma^2(C)=\sigma^2_{stat}+\sigma^2_{dyn}=k$ which is a measure of cluster
multiplicity, if all the associated particles in the cluster are included in
one bin. In practice not all the associated particles in cluster can be
included in the same bin, and  $\sigma^2(C)$ can be regarded as
an ``effective'' cluster multiplicity. Thus the short range correlations lead
to positive dynamic fluctuations and increase  $\sigma^2(C)$ from 1. There
is also the  case where the short range correlations show long range effect
in $\sigma^2(C)$, if the correlated particles are selected into different but
close bins.

For  $\sigma(C)$ calculated from reconstructed event multiplicity,
it also includes additional detector effects. Since both the $C$ distribution
of the particles from the model event generator and the $C$ distribution of the
reconstructed multiplicity after detector response simulation are very close
to Gaussian distributions, we may treat the detector effects as Gaussian
response smearing and write the reconstructed multiplicity fluctuations 
$\sigma^2(C)$ as the sum of three parts: 
\begin{equation}
\sigma^2(C) = \sigma^2_{stat} + \sigma^2_{dyn} +\sigma^2_{det},
\end{equation}
where the $\sigma^2_{det}$ represents the detector effects.
Using the detector simulation to estimate $\sigma^2_{det}$, the dynamic
fluctuations can be extracted from the following simple
equation:
\begin{equation}
\sigma^2_{dyn}=\sigma^2(C)-\sigma^2_{stat}-\sigma^2_{det}.
\end{equation}


This analysis used data taken by the PHOBOS 
detectors\cite{phobos_det} during the runs providing Au+Au collisions 
at $\sqrt{s_{NN}}=200$GeV. Only those data taken with a zero magnetic 
field setting and minimum bias triggers were used. The PHOBOS octagonal 
multiplicity detector was used to measure the charged particle 
multiplicity. It consists of 92 silicon sensors which surround the beam 
pipe and cover the pseudo-rapidity range of $|\eta|<3$. The active 
elements of the octagonal detector are constructed of highly segmented 
silicon pads which yield signals at the passage of charged 
particles. Within this detector, every second
azimuthal octant  has a window for another detector (spectrometer or vertex
detector), which leads to incomplete azimuthal coverage.
To reduce the systematic uncertainties in the fluctuations results,
these nonuniform octants were excluded in the analysis of the reconstructed 
multiplicities from both the data and simulations,  which left a 
total nominal azimuthal coverage of 50\%.  

In each $\eta$ bin, the sum of the angle-normalized energy deposits ($dE/dx$) 
was used to estimate the multiplicity of charged particles in a collision 
event. Assuming that the 
number of particles hitting a silicon pad follows  a Poisson  distribution, the number 
of recorded hits ($N_{hit}$) in  the $\eta$ bin can be fitted as a
function of the sum of the normalized energy deposits($E$) 
$N_{hit} = N_{max}(1-e^{-E/E_{max}})$ to extract two parameters $N_{max}$  
and $E_{max}$.  The parameter $N_{max}$ can be regarded 
as the effective number of active pads in the investigated $\eta$ bin and 
$E_{max}/N_{max}$ is then taken as the average energy deposited by one 
charged particle in this $\eta$ bin. Thus the multiplicity in the
investigated bin can be estimated as $N = E/(E_{max}/N_{max})$. 
From estimated charge multiplicities in two symmetric bins, $N_{1}$ and $N_{2}$, the variable $C$ 
is calculated for each event. The $\eta$ bins are chosen to have equal 
size and to be symmetric about $\eta$=0 in order to assure similar 
multiplicities in forward and backward regions. The $RMS$ of the $C$ 
distribution in each centrality and pseudo-rapidity bin is calculated 
directly as the fluctuations variable $\sigma(C)$. 

\begin{figure}
  \centering
  \includegraphics[width=14cm]{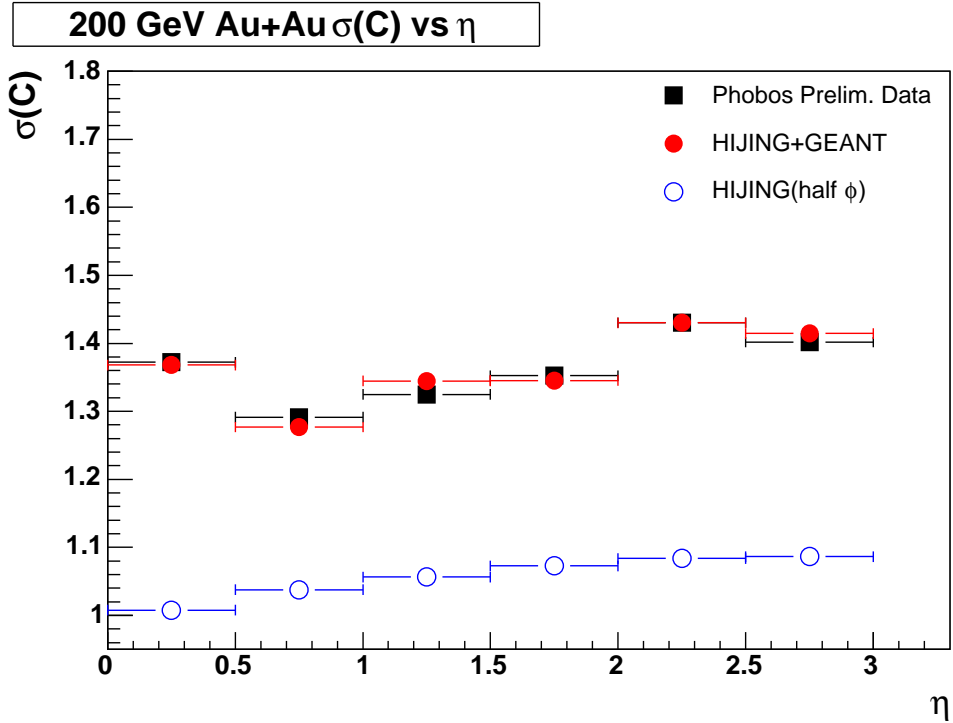}
  \caption{\label{fig:zwchai_fig1} 
$\sigma(C)$ versus $\eta$ with fixed $\eta$ bin size 0.5. 
All values of $\sigma(C)$ were calculated using $50\%$ azimuthal acceptance.
The estimated systematic error of $5\%$ is not shown in the plot.}
\end{figure}
\begin{figure}
  \centering
  \includegraphics[width=14cm]{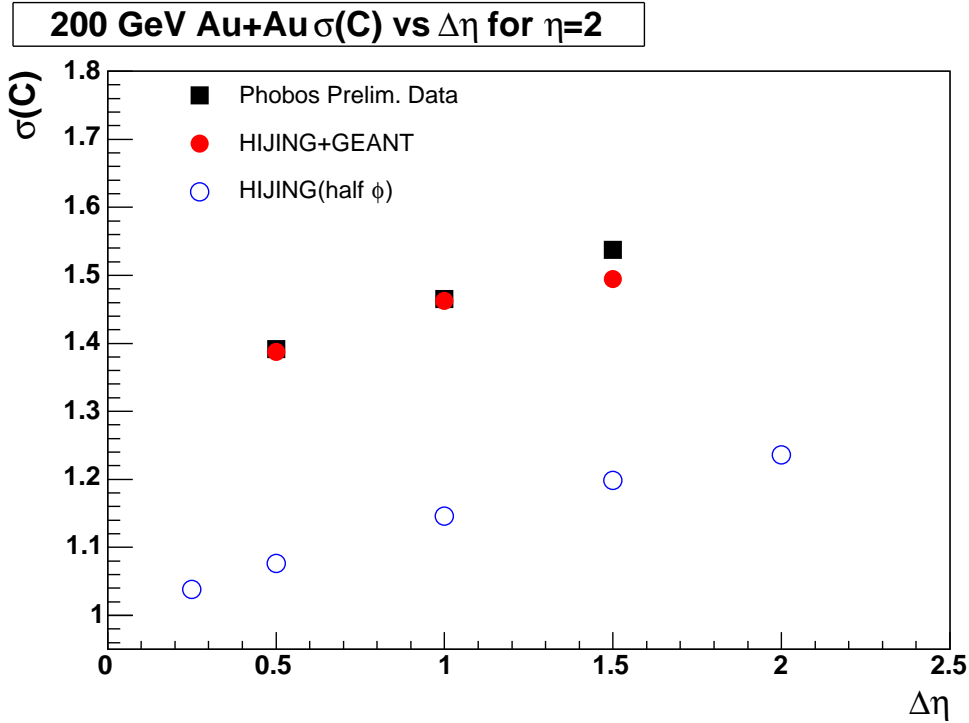}
  \caption{\label{fig:zwchai_fig2} 
$\sigma(C)$ versus $\Delta\eta$ with $\eta$ bin centered at 2.0. 
All values of $\sigma(C)$ were calculated using $50\%$ azimuthal acceptance.
The estimated systematic error of $5\%$ is not shown in the plot.}
\end{figure}
Some preliminary $\sigma(C)$ results\cite{qm04} are shown in Figure 1 and 2. 
Figure 1 shows the $\sigma(C)$ dependence on the separation in $\eta$ space with
fixed $\eta$ bin size 0.5. The data and reconstructed simulation
(HIJING\cite{hijing}+GEANT\cite{geant}) are similar. Both  have
a non-monotonic dependence on $\eta$ because of the presence of the detector
effects. The HIJING $\sigma(C)$ are slightly greater than 1, with the
difference between the $\sigma(C)$ and 1 increasing with the $\eta$
separation. This indicates that the dynamic fluctuations of HIJING are greater
than 0. The suppression in the mid-rapidity region is due to the fact that 
negative long range effect of the short range correlated particles selected into different
bins cancels the positive short range correlations effect. Figure 2 shows
the $\sigma(C)$ dependence on the $\eta$ bin size with the $\eta$ centered
at 2.0. The data and reconstructed simulation(HIJING+GEANT) are again
similar. The HIJING $\sigma(C)$ increases with the $\eta$ bin size. Such 
dependence on the bin size is a typical feature of particle correlations 
with finite correlation length. The data and reconstructed simulation 
$\sigma(C)$ also have the same trend of increasing with $\eta$ bin size.


Since detector effects dominate in the
reconstructed multiplicity fluctuations as shown in
Figure 1 and 2, it is necessary to clearly understand such effects 
in order to extract the dynamic fluctuations from the experimental
measurement. Extensive GEANT detector simulations with different
settings were used to investigate the relative importance of various
sources in the observed detector effects. Four main detector
effects were identified. 
The acceptance gap effect is significant in mid-rapidity region, 
but negligible in higher $\eta$ region.
Such acceptance gap effect in $\sigma(C)$ was caused by the 
asymmetric acceptance of the dead channels and gaps in eta coupled 
with the finite extension of the event vertex position.  
Since the energy deposit was
used in the evaluation of the charged particle multiplicity,
$dE/dx$ fluctuations also contribute to $\sigma_{det}^2$. The
$dE/dx$ variations are due to Landau fluctuations and the wide
range of detected particle velocities ($\beta$ variations). The
$\beta$ variation contribution is more significant in the mid-rapidity
region than in the high $\eta$ region. The Landau fluctuations
contribution is nearly constant throughout $\eta$ range covered by the 
octagonal 
detector. The contribution of secondary particles originating from primary
particle decays and interactions is important primarily in the
high $\eta$ region because of the longer path length and higher
probability of decay or interaction.


To extract reliable dynamic fluctuations from the data, the analysis
procedure was refined to suppress the detector effects as much as possible.
For those residual detector effects which can not be suppressed through
the improved analysis procedure, the simulations were used to reproduce them.
In the mid-rapidity region and central collisions, the acceptance gap effect
is the dominant detector effect. Through careful analysis of the cause of 
such detector effect, we designed an effective procedure for the suppression 
of the acceptance gap-induced detector effects by subtracting an 
event-by-event  $C$ offset, which is the mean of the $C$ distribution in 
each ($Z_{vtx}$, $\eta$, Centrality bin). This $C$ offset does not change the fluctuations due
to other sources.  Another improvement in the analysis is to use
$\eta$-dependent $dE/dx$ hit cuts to suppress detector effects from 
the secondary contribution in high eta region. Using the improved analysis
procedure, the detector effects in $\sigma(C)$ are greatly reduced and
significantly less than those shown in preliminary results in Figure 1 and
Figure 2. 

Three different simulations were used to evaluate the detector effects for
the extraction of dynamic fluctuations: standard HIJING+GEANT, standard 
AMPT\cite{ampt}+GEANT and modified HIJING+GEANT. The modified HIJING uses
the HIJING event generator to generate events and randomizes the sign in
$\eta$ of the particles, thus destroying correlations between the particles while keeping
other physics properties unchanged.  Equation (2) was used to extract the
detector effects from these simulations. The extracted detector effects are 
similar but have some small model dependence. Careful examination of the 
extracted detector effects' dependence on the model dynamic fluctuations
helped identifying a linear relation between the extracted detector effects 
and the dynamic fluctuations of the model used in detector simulation, which
is parametrized as
\begin{equation}
\sigma^2_{det}=\sigma^2_{det0}(1-\alpha\sigma^2_{dyn}),
\end{equation}
where the two parameters $\sigma^2_{det0}$ and $\alpha$ were estimated by
fitting the $\sigma^2_{det}$ as a function of $\sigma^2_{dyn}$ 
for each $\eta$ bin. Such function fit was possible since there were three
different simulations. The fitted  $\sigma^2_{det0}$ and $\alpha$ were then
used to construct adjusted detector effects for the extraction of dynamic 
fluctuations with the following equation:
\begin{equation}
\sigma^2(C)=\sigma^2_{stat}+ \sigma^2_{dyn} +
\sigma^2_{det0}(1-\alpha\sigma^2_{dyn}).
\end{equation}

Since the multiplicity was calculated using only half of the azimuthal
acceptance, the extracted dynamic fluctuations were multiplied by the factor
2 to extrapolate dynamic fluctuations in half acceptance to that
corresponding to full azimuthal acceptance. Such extrapolation was validated
by the analysis of the dynamic fluctuations in simulations. In both HIJING 
and AMPT, the ratio of dynamic fluctuations for full azimuthal acceptance and dynamic
fluctuations for half azimuthal acceptance in all the investigated $\eta$ bins is 2.

\begin{figure}
  \centering
  \includegraphics[width=14cm]{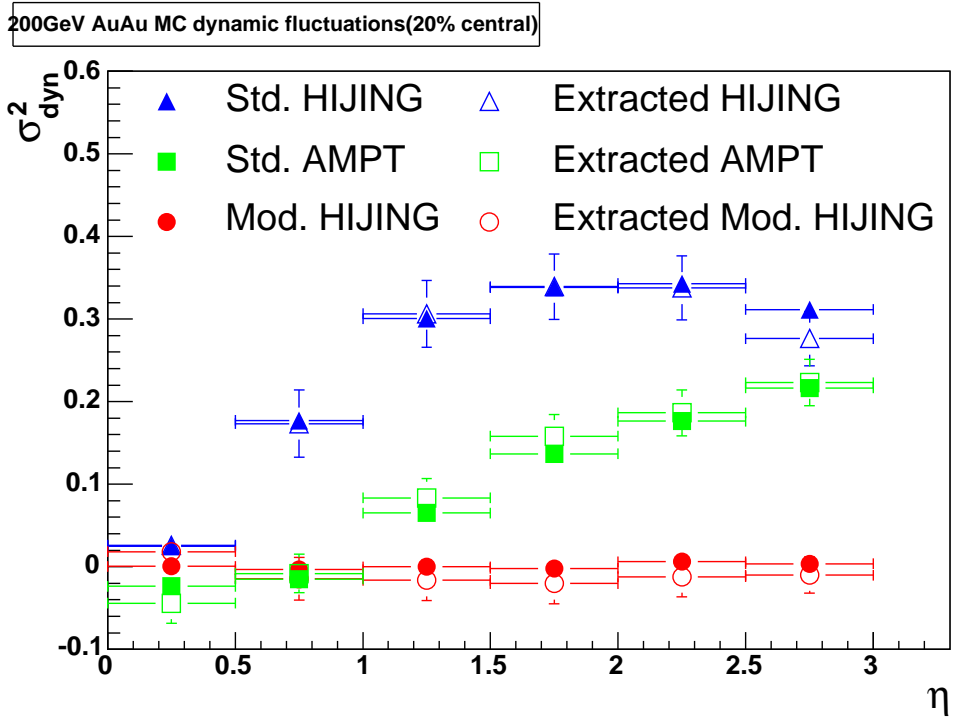}
  \caption{\label{fig:zwchai_fig3}
Comparison of extracted and true model dynamic fluctuations in  0-20%
central collisions. The dynamic fluctuations analysis procedure as 
described in the text was used to extract the dynamic fluctuations 
from the reconstructed multiplicity of the simulations.}
\end{figure}
The dynamic fluctuations extraction procedure was applied to the
reconstructed simulations to extract their dynamic fluctuations. As shown in
Figure 3, the extracted dynamic fluctuations are consistent with the true 
dynamic fluctuations in all three simulations. 
Thus the analysis procedure for the extraction of dynamic fluctuations from 
reconstructed multiplicity was validated by simulations. We are now applying 
this procedure to extract the dynamic fluctuations from the Au+Au data.
The analysis results compared with model predictions will be available in 
a later paper. We expect the new dynamic fluctuations results will provide 
useful insights on the intrinsic particle correlations in Au+Au collisions.


In summary, we analyzed the multiplicity fluctuations in the
inclusive charged particle production in Au+Au collisions at
$\sqrt{s_{NN}}=200$GeV within the pseudo-rapidity range  of
$-3<\eta<3$.
The preliminary fluctuations in the data are similar to that in
reconstructed simulations(HIJING+GEANT). The HIJING multiplicity 
fluctuations grow with increasing $\eta$ bin separation and width, 
revealing the existence of particle correlations in Au+Au collisions.
An improved analysis procedure for extracting the dynamic fluctuations 
from the data was developed. This procedure was validated with 
simulations and it is now being used to extract the dynamic fluctuations 
from experimental data. The new data results will be available in near
future. We expect these results will provide helpful insights on 
the intrinsic correlations in particle production. They could also be 
very useful input for theoretical model development for understating 
the particle production mechanism in heavy ion collisions.

{\large\bf Acknowledgments} \\
%
%
%
%
This work was partially supported by: U.S. DOE grants 
DE-AC02-98CH10886,
DE-FG02-93ER40802, 
DE-FC02-94ER40818,  
DE-FG02-94ER40865, 
DE-FG02-99ER41099, and
W-31-109-ENG-38, 
NSF grants 9603486, 
0072204,            
and 0245011,        
Polish KBN grant 2-P03B-10323, NSC of Taiwan contract NSC 89-2112-M-008-024.


\begin{thebibliography}{99}

\bibitem{phase_fluc} Stephanov M, Rajagopal K  and Shuryak E, Phys. Rev. {\bf D60} (1999) 114028
\bibitem{ua5} Alpgard K {\it et al.}, Phys. Lett. {\bf B123} (1983) 361
\bibitem{phobos_det} Back B B {\it et al.}, (PHOBOS), Nucl. Instrum. Methods Phys. Res.,
Sect. A {\bf 499} (2003) 603
\bibitem{qm04} Wozniak K {\it et al.}, (PHOBOS), J. Phys. {\bf G30} (2004) S1377-S1380
\bibitem{hijing} Gyulassy M and Wang X N, Phys. Rev. {\bf D44} (1991) 3501
\bibitem{geant} Brun R {\it et al.}, GEANT 3.21, Detector Description and Simulation
Tool, CERN Program Library Long Write-up W5013, 1994
\bibitem{ampt} Zhang B and Kuo C M and Li B A and Lin Z, Phys. Rev. {\bf C61} (2000) 067901

\end{thebibliography}
\end{document}